\documentclass[aps,prl,preprint,showpacs,superscriptaddress,groupedaddress]{revtex4-1}  

\usepackage{graphicx}  
\usepackage{dcolumn}   
\usepackage{bm}        
\usepackage{amssymb}   

\begin{document}


\title{Boosting Room-Temperature Li$^+$ Conductivity via strain in solid electrolytes for Lithium-ion Batteries}
\author{Rodolpho Mouta}
\affiliation{Departamento de F\'{\i}sica, Universidade Federal do Maranh\~ao, S\~ao Luis, Maranh\~ao, 65080-805, Brazil}
\author{Carlos William de Araujo Paschoal}
\affiliation{Departamento de F\'{\i}sica, Universidade Federal do Cear\'a, Fortaleza, Cear\'a, 60455-554, Brazil}
\email{paschoal.william@fisica.ufc.br}
\date{\today}

\begin{abstract}
Room-temperature (RT) conductivity of most candidates for solid electrolytes of miniaturized lithium-ion batteries is still 1-2 orders of magnitude below commercial requirements, therefore several approaches are being pursued aiming the enhancement of such values. In this letter, we report for the first time direct theoretical evidence that epitaxial strain can be an efficient tool to reach this goal.
\end{abstract}

\pacs{95.55.Sh, 93.90.+y, 13.15.+g}
\maketitle

Significant effort is now being devoted to developing solid materials that can replace Li-based liquid electrolytes currently used in lithium-ion batteries (LIBs), since Li-based liquid electrolytes are flammable, toxic, incompatible with the highest specific energy anode (lithium metal), and pose a limit to miniaturization.\cite{Quartarone2011,Choi2012,Zhao2012b,Lu2016} However, at room-temperature (RT), Li$^+$ conductivity in the best solid electrolytes ($\leq 25$ mS/cm) is still 1-2 orders of magnitude below that in liquid counterparts ($>100$ mS/cm).\cite{Kamaya2011,Braga2016,Seino2014,Tang2015} Several attempts are being performed to solve this issue, including search and design of new materials as well as the properties tailoring of well-established materials using non-stoichiometry, doping, chemical substitutions and refined synthesis techniques. \cite{Zhao2012b,Kamaya2011,Braga2016,Seino2014,Tang2015,Sahu2014,Deng2015} An approach that has been proven particularly efficient to improve O$^{2-}$ transport in electrolytes of miniaturized all-solid-state fuel cells is the use of epitaxial strain obtained by depositing a thin-film on a substrate with a mismatching lattice parameter.\cite{Garcia-Barriocanal2008,Sillassen2010,DeSouza2012,Kubicek2013,Yildiz2014,Goyal2015,Lee2015} Recently, Tealdi et al \cite{Tealdi2016} also used this approach to predict a one order of magnitude enhancement at 500 K for cathodes of Li- and Na-ion batteries, with an even larger enhancement being expected at room temperature. Notwithstanding, no analogue strain-induced enhancement has been reported so far in Li$^+$-conducting candidates for solid electrolytes in miniaturized LIBs. \cite{Bachman2016,Wei2015}

In this Letter, we show clearly a significant Li$^+$ conductivity enhancement (by orders of magnitude) induced by strain in a solid electrolyte, opening a new window for ionic transport improvement in such materials, aiming their use in miniaturized LIBs. Our result was obtained by calculating an analytical expression for the relative change that strain induces in Li$_3$OCl, a prototype member of a promising family of potential solid electrolytes for LIBs that has received increasing attention in the past few years: the so-called Lithium-Rich Anti-Perovskites (LiRAPs). \cite{Zhao2012b,Lu2016,Braga2014a,Deng2015,Reckeweg2012,Zhang2013d,Emly2013a,Mouta2014,Zhang2014a,Lu2014a,Schroeder2014,Zhang2015,Braga2016,Chen2015,Lu2015,Li2016,Deng2016,Gao2016,Mouta2016} Such expression was obtained in terms of the Gibbs energy barriers for Li$^+$ migration, which were evaluated numerically by computational modeling.

{\it Analytical model and numeric evaluation - } The analytical expression for ionic conductivity $\sigma$ was calculated according to the hopping model, which can always be used as long as one knows the crystal structure, all relevant charge carriers, their concentration and the possible migration paths. Thus, considering the conductivity as \cite{Bishop2013}
\begin{equation} \label{1}
\sigma = \sum_i n_i \left|q_i\right| \mu_i,
\end{equation}
we used statistical thermodynamics to calculate the concentration $n_i$ of each one of the charge carriers, whose charges are $q_i$, lying at distinct positions $i$ inside the unit cell at a temperature $T$. The mobility $\mu_i$ of each of them is given by \cite{Bishop2013,Lidiard1957}
\begin{equation}\label{mu_i}\label{2}
  \mu_i = \frac{\left| q_i \right| \nu \beta}{2k_BT} \sum_j \left(\rho_{i\rightarrow j} \right)^2 \exp \left(-{G_{i\rightarrow j}^{(m)}/k_BT}\right),
\end{equation}
in which we take into account all neighboring positions $j$ to which the carrier can migrate, these lying at distances (projected at the external applied electric field direction) $\left| \rho_{i\rightarrow j} \right|$ from position $i$. The above expression for mobility is based on the fact that between positions $i$ and $j$ there is a Gibbs energy migration barrier $G_{i\rightarrow j}^{(m)}$ that the charge carrier tries to “hop” with attempt frequency $\nu$, having probability of success at each try given by the exponential term. We highlight that this model has been widely and successfully used both by theoreticians and experimentalists to replicate, model and explain the ionic transport in crystalline materials. \cite{Bishop2013,Lynch1960,Fuller1968,Fuller1968a,PhysRevB.11.1654,Dorenbos1987,Tschope2001,Catlow1979}

For a unstrained cubic (anti-)perovskite-structured ABX$_3$ crystal having vacancies of X ion as charge carriers, the ionic conductivity can be written as \cite{Nowick1999}
\begin{equation}\label{sigma0}\label{3}
  \sigma_o = \frac{2}{3} \nu \beta n \left( a_o \left| q_X \right| \right)^2 \exp \left( -\beta G_o^{(m)} \right),
\end{equation}
in which $\beta = \left( k_B T \right)^{-1}$ with $k_B$ being the Boltzmann constant, $a_o$ is the unstrained lattice parameter, $G_o^{(m)}$ is the migration Gibbs energy from one vertex of the BX$_6$ octahedron to another and $n$ is the amount of X vacancies per unit volume. In doped or non-stoichiometric crystals this amount usually is strain- and temperature-independent, being defined by the doping or non-stoichiometry degree, as long as the concentration of thermally activated charge carriers can be neglected, as usually happens at RT. This is, for example, the case of Li$_3$OCl antiperovskite,  in which the number of thermally activated charge carriers per unit formula is $\sim 10^{-7}$ \cite{Mouta2014}, while the corresponding values due to doping and non-stoichiometry are $> 5 \times 10^{-3}$ \cite{Braga2014a,Zhang2014a} and $\sim 10^{-2}$ \cite{Zhao2012b}, respectively.

However, if a biaxial stress is applied in the [100] and [010] directions, cubic symmetry is broken and Eq.  (\ref{3}) no longer holds. Considering that no octahedral tilt occurs and all ions remain in their original fractional coordinates relative to the lattice parameters, the basal plane contraction (compressive stress) or expansion (tensile stress) leads to a tetragonal structure. Thus, the X ions in planes (001)(in apical positions relative to the BX$_6$ octahedra) and X ions in planes (002) (in equatorial positions relative to the octahedra) are no longer equivalent by symmetry. This implies that apical and equatorial vacancies will have distinct formation Gibbs energies $G_{ap}^{(\nu)}$  and $G_{eq}^{(v)}$, so that ion X vacancies will tend to concentrate in planes with lowest Gibbs energy. Also, there are now three distinct vacancy migration paths: (i) equatorial-to-equatorial, (ii) equatorial-to-apical, and (iii) apical-to-equatorial, with respective migration Gibbs energies $G_{eq \rightarrow eq}^{(m)}$, $G_{eq\rightarrow ap}^{(m)}$ and $G_{ap\rightarrow eq}^{(m)}$ (apical-to-apical migration barrier is always large, with negligible contribution to mobility).

This leads to distinction between lateral (i.e., in the same plane of applied stress) and perpendicular (to the plane of applied stress) conductivities. We noticed strain was always detrimental to perpendicular transport in Li$_3$OCl (and apparently for perovskites in general), so here we focus on the lateral conductivity only, which on this case is
\begin{equation}\label{sigma_2}\label{4}
  \sigma = \nu \beta n \left( a \left | q_X \right| \right )^2 e^{-\beta G_{eq \rightarrow eq}^{(m)}} \left[ \frac{1+e^{-\beta(G_{eq\rightarrow ap}^{(m)}-G_{eq\rightarrow eq}^{(m)})}} {2+e^{-\beta(G_{ap}^{(\nu)}-G_{eq}^{(\nu)})}} \right],
\end{equation}
where $a$ is the strained lattice parameter parallel to any one of the two equivalent directions of applied stress. Thus the ratio between lateral ionic conductivities of the strained and the unstrained crystal is
\begin{equation}\label{sigma_3}\label{5}
  \frac{\sigma}{\sigma_o} = \frac{3}{2} \left( \frac{a}{a_o} \right)^2 e^{\beta \left( G_o^{(m)}-G_{eq \rightarrow eq}^{(m)} \right)} \left[ \frac{1+e^{-\beta (G_{eq\rightarrow ap}^{(m)}-G_{eq\rightarrow eq}^{(m)})}} {2+e^{-\beta(G_{ap}^{(\nu)}-G_{eq}^{(\nu)})}} \right]
\end{equation}

Therefore, to determine the relative increase (or decrease) in Li$_3$OCl's Li$^+$ conductivity as a function of biaxial strain, we just need to know how the lattice parameter, as well as Gibbs energies of formation and migration of Li$^+$ vacancies, depend on strain. Classical atomistic modelling is a powerful and reliable tool to access defect energetics in energy materials (including LIB materials) \cite{Mouta2014,Mouta2016,Schroeder2013a,Panchmatia2014a,Islam2014a,Clark2014a,Deng2015a,Tealdi2016} and to estimate the effect of strain on ionic transport.\cite{DeSouza2012,Tealdi2016,Dezanneau2012,Rushton2014,Tealdi2014} Thus, we performed quasi-static calculations based on pair-wise interionic potentials using the GULP code  \cite{A606455H,Gale2003,Gale2005}  to obtain the lattice parameter and Li$^+$ vacancies formation and migration Gibbs energies required in Eq.  (\ref{5}) as a function of tensile biaxial strain for Li$_3$OCl. We adopted a robust, already validated parameterization previously derived by our group,\cite{Mouta2014,Lu2015,Mouta2016} since the resulting force field has the unusual advantage of allowing the straightforward calculation of Gibbs energies, instead of enthalpies. This is important because neglecting entropy contributions may lead to uncertainties as large as one order of magnitude.\cite{DeSouza2012} All parameters in Eq.  (\ref{5}) were calculated at 300 K, since this is the ideal LIB operation temperature. We also point out that in the range of biaxial strains investigated (0-6\%), the only structural change observed in our calculations was a simple tetragonal distortion, without any octahedral tilts, so that all assumptions leading to Eq.  (\ref{5}) were satisfied.

\begin{figure}[ht]
\includegraphics[scale=0.4]{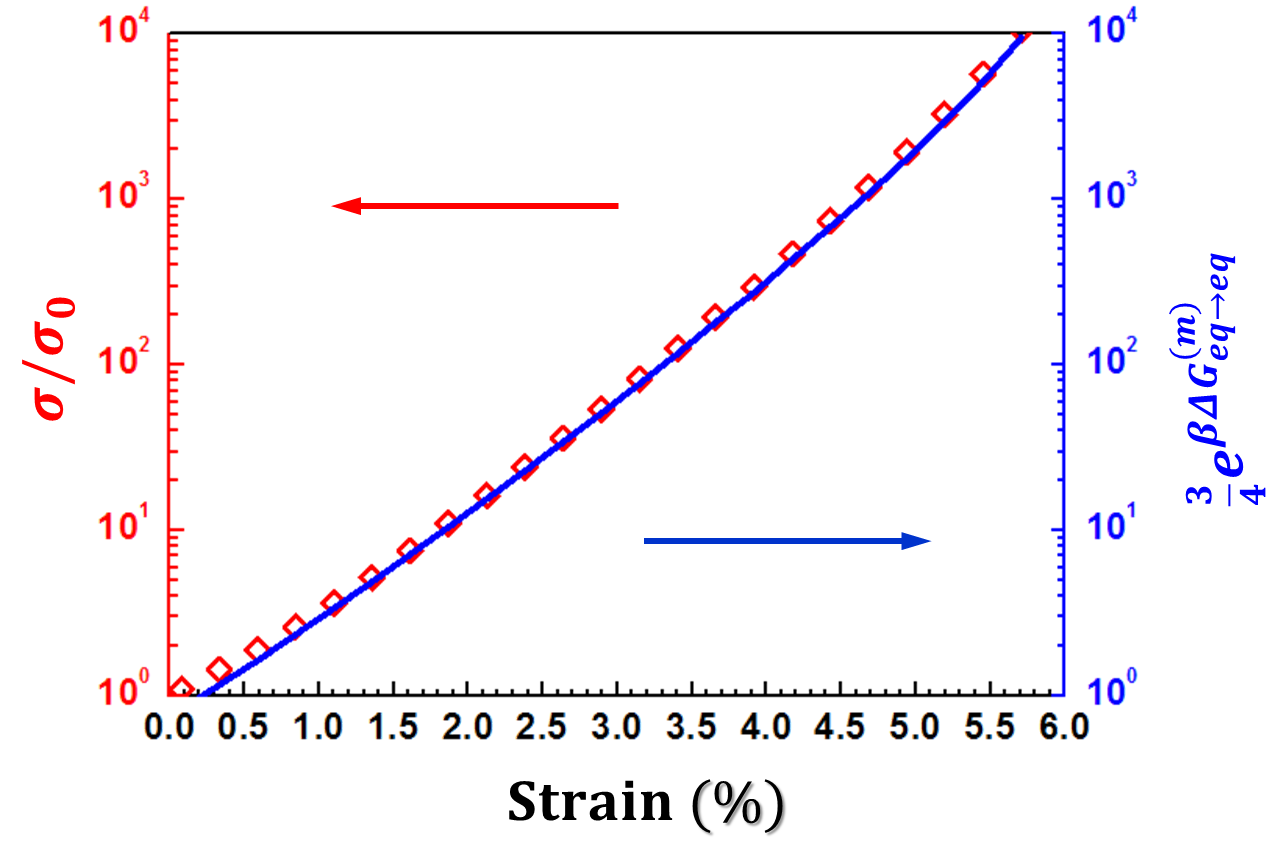}
\caption{\label{fig1} (Color online) Ratio between lateral conductivities of strained and unstrained Li$_3$OCl for a tensile biaxial strain in the directions [100] and [010], calculated from both the full expression, Eq.  (5) (red diamonds), and the simple approximation given by Eq.  (9) (blue line).}
\end{figure}

{\it Ionic conductivity boost and implications for epitaxial films - } The calculated relative enhancement in lateral Li$^+$ conductivity as a function of tensile biaxial strain, as shown in Fig. \ref{fig1}, indicates that a huge conductivity increase, nearly with exponential behavior, can be achieved with relatively low strains, with the ionic conductivity increasing by over one order of magnitude at 2\% and reaching a two orders of magnitude enhancement already at 3.3\%. Considering that the RT Li$^+$ conductivities of Li$_3$OCl-based polycrystals $(10^{-4}-10^{-3}$ S/cm$)$ \cite{Zhao2012b,Kamaya2011,Braga2014a,Lu2014a} are among the highest ones reported for crystalline materials, the possibility of enhancing such values by even 1-2 orders of magnitude is remarkable, since this may improve significantly this electrolyte performance in miniaturized LIBs. We stress that this is the first clear evidence of strain-induced enhancement in ionic transport of a Li$^+$-conducting solid electrolyte \cite{Bachman2016}. For completeness sake, we also investigated influence of tensile biaxial strain over perpendicular conductivity (not shown here), but it was found to be detrimental. Also, further calculations for compressive biaxial strains revealed that these are detrimental not only to perpendicular, but also to lateral transport. A detailed investigation of Li$_3$OCl's ionic conductivity dependence on different strain states is beyond the scope of this Letter and will be published elsewhere.

These findings have important implications for Li$_3$OCl thin films use as solid electrolytes in miniaturized LIBs. The commonest way to obtain a tensile strain is by depositing an epitaxial film on a substrate whose lattice parameter is larger than that of the ionic conductor. For Li$_3$OCl, a commercially  available substrate which implies a tensile strain around 2$\%$ is K(Ta$_{1-x}$Nb$_x$)O$_3$ (KTN), whose lattice parameter ranges from 3.989 to 4.000 \AA {} depending on Nb concentration. Commonly used substrates as SrTiO$_3$, LaAlO$_3$ and YAlO$_3$ have lattice parameter lower than Li$_3$OCl and should not imply in a conductivity enhancement if Li$_3$OCl grows in [001] direction on these substrates. Also, substrates  with very larger lattice parameter which are almost $a_o$ multiples can be used, but in this case the epitaxial deposition is difficult. Finally, a non-strained thick film with the desired lattice parameter can be grown in a conventional substrate as an intermediated layer to drive a tensile strain in Li$_3$OCl. For example, Li$_3$OBr ($a \sim 4.045$ \AA {}) can be used as such intermediate layer, delivering a tensile strain of {\it ca.} 3.5\%.

\begin{figure}[ht]
\includegraphics[scale=0.4]{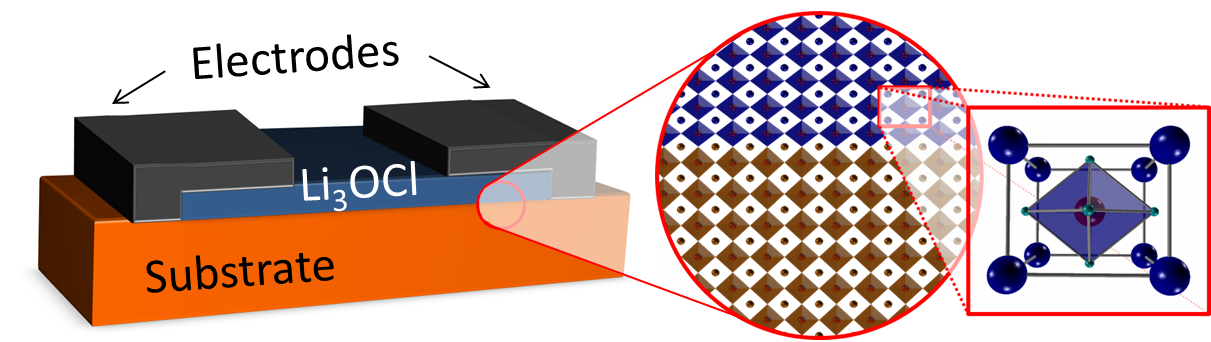}
\caption{\label{fig2} (Color online) Apparatus suggested for future experimental probe of lateral conductivity enhancement in a Li$_3$OCl thin film, induced by epitaxial strain due to lattice parameter mismatch between the film and substrate. The inset shows the resulting tensile strain in the Li$_3$OCl's unit cell.}
\end{figure}

Furthermore, the electrodes should be placed on the sides of the film (see Fig. \ref{fig2}), instead of sandwiching it from below and above, as it is usually performed for perovskites' epitaxial growth by sputtering or pulsed laser deposition (PLD), when a conductor perovskite layer, as SrRuO$_3$, is first deposited. It is necessary to use lateral electrodes so that lateral, instead of perpendicular conductivity, is obtained. This setup has already been successfully employed to observe a two orders of magnitude enhancement in lateral ionic conductivity at 573 K due to tensile biaxial strain in a yttria-stabilized zirconia (YSZ) thin film deposited on SrTiO$_3$ (STO) substrate. \cite{Sillassen2010} Another possible setup is to place both electrodes on top, instead of on the sides. This is similar to the setup employed by Garc\'{\i}a-Barriocanal {\it et al.} \cite{Garcia-Barriocanal2008} to observe an eight orders of magnitude enhancement in lateral conductivity of YSZ near RT (357 K) in YSZ/STO heterostrucures, which has been partially attributed to strain.\cite{DeSouza2012,Pennycook2010} However, it has been argued that this later setup may lead to ``enormous electrode resistance" \cite{Guo2009} due to the small contact between the electrodes and the system under study, so that the one presented in Fig. \ref{fig2} may be more suitable.

\begin{figure*}[ht]
\includegraphics[scale=0.35]{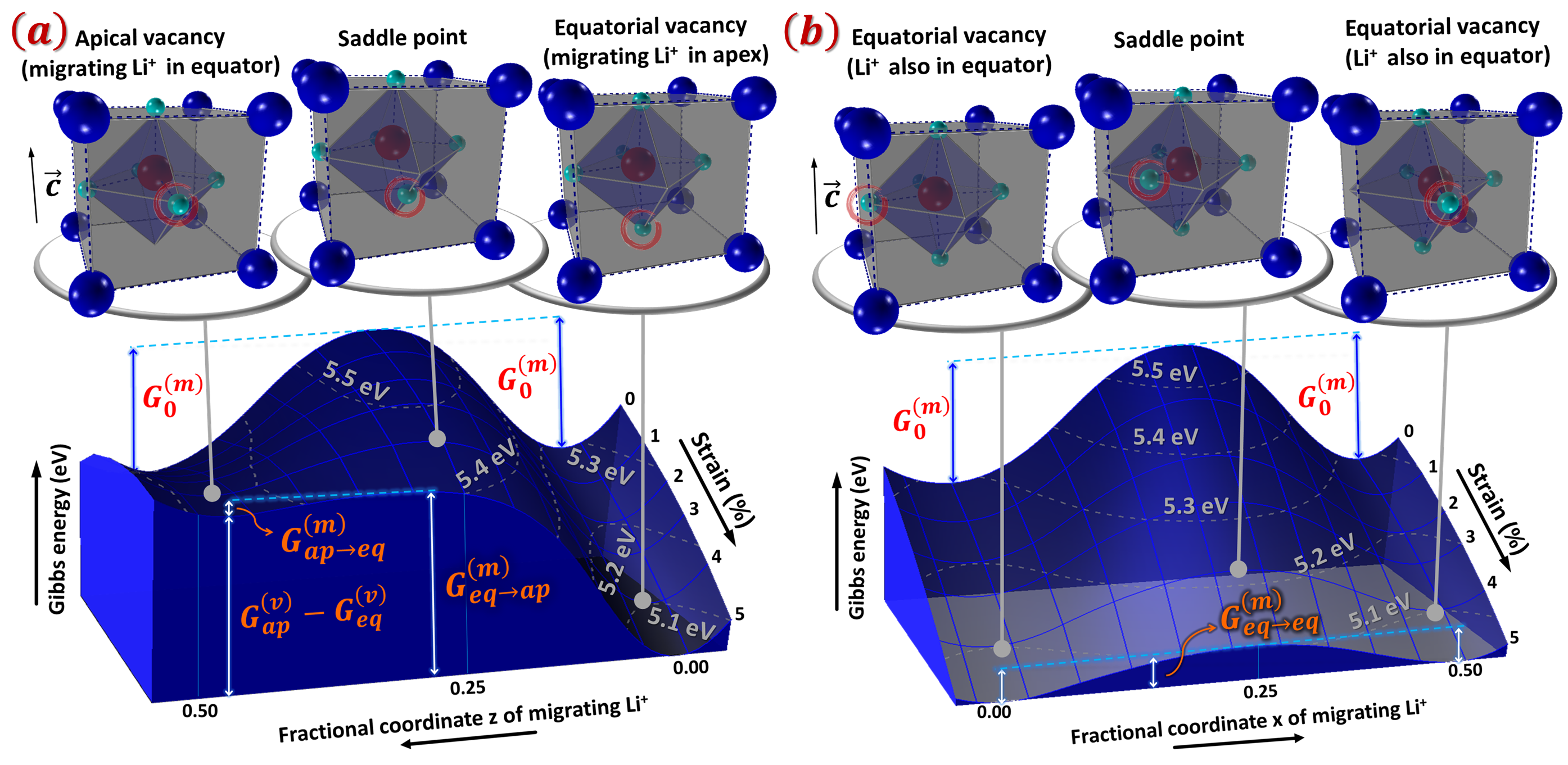}
\caption{\label{fig3}(Color online) Gibbs energy barrier profiles as a function of strain for (a) equatorial-to-apical and (b) equatorial-to-equatorial migration, obtained from atomistic modelling. The grey dashed contours provide reference to numeric values of Gibbs energy, while blue filled lines evidence the barrier profile for a fixed strain value. Superior insets show snapshots of the relaxed structure around the migrating Li$^+$ vacancy, with Li$^+$, O2- and Cl- colored cyan, red and blue, respectively, and blue dashed lined connecting ions to help visualizing structural distortions in relation to the initial structure (i.e., undistorted  and without vacancies).}
\end{figure*}

{\it Exponential dependence of ionic conductivity on strain - } Along with the conductivity enhancement in itself, its simple exponential dependence on strain is also very interesting from a physical point of view, once it is not directly obvious from Eq.  (\ref{5}). Therefore, we further investigated this dependence, getting helpful physical insights along the way, as shown next. For tensile strains, the calculated Gibbs energy of formation of equatorial vacancies was lower than of apical vacancies, as can be seen in Fig. \ref{fig3}(a), so that
\begin{equation}
G_{ap}^{(\nu)}-G_{eq}^{(\nu)} > 0.
\end{equation}
Also, according to Fig. \ref{fig3}(b), the Gibbs energy barrier for equatorial-to-equatorial migration is lower than the unstrained counterpart. Thus, for non-zero strain,
\begin{equation}
\Delta G_{eq\rightarrow eq}^{(m)} \equiv G_0^{(m)} - G_{eq \rightarrow eq}^{(m)} > 0.
\end{equation}

As also shown in Fig. 3(a), the equatorial-to-apical migration barrier is higher than the unstrained one, since Gibbs energy of formation of an equatorial vacancy decreases faster with strain than the saddle point Gibbs energy does. For example, for 5\%, the contours show that Gibbs energy of formation of an equatorial vacancy is reduced by more than 0.2 eV, while the saddle point Gibbs energy is reduced by less than 0.2 eV. Accordingly, also for non-zero strain,
\begin{equation}\label{G1}
G_{eq\rightarrow ap}^{(m)}- G_{eq\rightarrow eq}^{(m)} > 0.
\end{equation}
Our calculations showed all these differences grow almost linearly when strain increases, in a rate high enough for the corresponding exponentials of the differences present in Eq. (6) and in Eq. (8) to become negligible for strains greater than ~1\%. In this case, Eq.  (\ref{5}) reduces simply to
\begin{equation}
\frac{\sigma}{\sigma_o} \approx \frac{3}{4} e^{\beta \Delta G_{eq\rightarrow eq}^{(m)}} ,	
\end{equation}
since the variation in the lattice parameter is small compared to the exponential above. Therefore, for strains greater than ~1\%, Eq.  (9) can provide a great estimate of relative increase in Li$_3$OCl's lateral ionic conductivity compared to the full expression, as observed in Fig.\ref{1}. Notice Eq.(9) directly explains the nearly exponential behavior obtained, since the dependence of equatorial-to-equatorial Gibbs energy barriers on strain was found to be almost linear.

Despite this exponential impact of strain over conductivity, it is  important to observe that strain may not be so helpful to increase conduction at high temperatures, once the relative increase in conductivity has exponential dependence also on the inverse of temperature. This is consistent with molecular dynamics calculations at very high temperatures on LaGaO$_3$, where the increase in lateral diffusion was small, yet exponential on strain.\cite{Tealdi2014} However, while performance at high/intermediate temperatures is critical for a potential solid oxide fuel cell electrolyte such as LaGaO$_3$, the ideal operation temperature of a LIB is RT, so the fact that conductivity enhancement at higher temperatures may be small is not a problem for Li$_3$OCl. Besides, its conductivity already increases rapidly to very high values at superior temperatures, due to its low activation enthalpy.

{\it Microscopic insights -} Now we turn our attention to the terms present in Eq. (5), but missing in Eq. (9), since this comparison leads to further insights about the the preferential migration path and the distribution of vacancies between apical and equatorial sites. First, we observe that the main feature contributing for the strain-induced enhancement in Li$^+$ conduction is the reduction of equatorial-to-equatorial barrier, without any significant influence from the other two. This can be explained by degeneracy breaking of vacancy formation Gibbs energy, which splits into two values, one for each plane. On one hand, this split makes the (001) planes energetically unfavorable for vacancies, so that too few vacancies (~2\% for strains as low as 1\%) are in apical positions for the apical-to-equatorial hopping contribution to be relevant. On the other hand, it increases Gibbs energy of equatorial-to-apical hopping, once the ground state for migration lowers (more than the saddle point does; see Fig. \ref{3}(a)), limiting significantly the amount of hops by this route. Therefore, lateral conduction happens almost exclusively in (002) planes; accordingly, only the migration barrier in these planes are relevant for strains $> \sim 1\%$.

Second, at first sight, the transport in (002) planes is not strongly influenced by the drastic increase of equatorial vacancy concentration, once Eq.  (9) has no dependence on the relative formation Gibbs energies. This is intriguing, since the conductivity is linearly proportional to the charge carriers concentration in each plane and should be very sensitive to them, as described by Eq.  (1). However, a closer look in the statistical distribution of vacancies over the two planes as a function of strain reveals that the percent of equatorial vacancies is already $\sim$98\%  for strains as low as 1\%, being $>$99\% when strain is 1.4\%, hence with only minor increments. Therefore, for strains greater than ~1\%, the concentration of equatorial vacancies saturates, reaching a practically constant value. Accordingly, its contribution to Eq.  (9) is also a constant value, embedded in the pre-exponential term.


{\it Further applicability -} We highlight that despite the considerations that led to Eq. (9) did not take into account octahedral tilts, apparently this exponential dependence is also valid when tilts are present, as in the case of perovskite LaGaO$_3$, for example. It is easy to verify that our very simple Eq. (9) is able to reproduce with remarkable accuracy the relative increase in LaGaO$_3$'s lateral transport as obtained by Tealdi and Mustarelli using molecular dynamics, \cite{Tealdi2014} an approach that is much more computationally expensive and time consuming than static defect calculations.

We infer that the applicability of Eq. (9) to (anti)perovskite-structured materials is primarily associated to the reduction in equatorial-to-equatorial migration barrier and to the strain-induced split of vacancy formation Gibbs energy, and hence reallocation of all vacancies to equatorial positions already for small strains, instead of other structural features such as octahedral tilts. More interestingly, such exponential dependence seems to hold even for structures other than perovskite, such as fluorite, as shown by De Souza {\it et al.} for ceria.\cite{DeSouza2012} Therefore, the exponential dependence appears to have a somewhat general validity and turns out to be a simple way to estimate the effect of biaxial strain on parallel conductivity from theoretical values of Gibbs energy barriers or enthalpy barriers, as an approximation to Gibbs energy when migration entropy does not vary significantly with strain.

{\it Conclusions and perspectives -} Summarizing, in this Letter we showed that strain is a powerful tool to boost RT Li$^+$ conductivity in thin-film solid electrolytes for miniaturized LIBs. We exemplified this by calculating the relative enhancement in ionic conductivity that biaxial tensile strain can induce in a promising solid electrolyte, Li$_3$OCl, in which a two orders of magnitude increase was predicted for a 3.3\% strain. We also provided practical suggestions on experimental setup and strategies to observe this effect, which may lead to the larger Li$^+$ conductivity ever reported in a solid electrolyte. We established an analytical expression for the strain-induced enhancement valid for vacancy migration between octahedra vertices in perovskite-structured solids in general, so it can also be used to probe the same effect in a wide range of other O$^{2-}$, Li$^+$ and Na$^+$ conductors candidates for solid electrolytes in fuel cells or alkali-metal-ion batteries, such as perovskite oxides, \cite{Tealdi2014} other LiRAPs and also NaRAPs (Na-Rich Anti-Perovskites) \cite{Wang2015a,Wang2016}. Finally, the approach used here can also be scaled for more complex structures other than perovskite.

We thank J. Gale for permitting us use the GULP code and the Brazilian funding agencies CAPES and CNPq for partially supporting this research. We are also grateful to E. F. V. Carvalho and E. N. Silva for computational facilities, and C. A. C. Feitosa for enlightening discussions.

\end{document}